\documentstyle[preprint,aps,eqsecnum]{revtex}
\tighten
\begin{document}
\preprint{PITT-93-6; LPTHE-93-52; CMU-HEP-93-21; DOR-ER/40682-46}
\draft
\title{\bf NON-EQUILIBRIUM EVOLUTION OF SCALAR FIELDS IN FRW COSMOLOGIES I}
\author{{\bf D. Boyanovsky$^{(a)}$, H.J. de Vega$^{(b)}$
 and R. Holman$^{(c)}$}}
\address
{ (a)  Department of Physics and Astronomy, University of
Pittsburgh, Pittsburgh, PA. 15260, U.S.A. \\
 (b)  Laboratoire de Physique Th\'eorique et Hautes Energies$^{[*]}$
Universit\'e Pierre et Marie Curie (Paris VI) ,
Tour 16, 1er. \'etage, 4, Place Jussieu
75252 Paris, Cedex 05, France \\
 (c) Department of Physics, Carnegie Mellon University, Pittsburgh,
PA. 15213, U. S. A.}
\date{10-19-93}
\maketitle
\begin{abstract}
We derive the effective equations for the out of equilibrium time evolution of
the order parameter and the fluctuations of a scalar field theory in spatially
flat FRW cosmologies.The calculation is performed both to one-loop and in a
non-perturbative, self-consistent Hartree approximation.The method consists of
evolving an initial functional thermal density matrix in time and is suitable
for studying phase transitions out of equilibrium.  The renormalization aspects
are studied in detail and we find that the counterterms depend on the initial
state. We investigate the high temperature expansion and show that it breaks
down at long times.  We also obtain the time evolution of the initial Boltzmann
distribution functions, and argue that to one-loop order or in the Hartree
approximation, the time evolved state is a ``squeezed'' state. We illustrate
the departure from thermal equilibrium by numerically studying the case of a
free massive scalar field in de Sitter and radiation dominated cosmologies. It
is found that a suitably defined non-equilibrium entropy per mode increases
linearly with comoving time in a de Sitter cosmology, whereas it is {\it not} a
monotonically increasing function in the  radiation dominated case.
\end{abstract}

\newpage

\section{\bf Introduction and Motivation}

Since its creation more than a decade ago, the inflationary universe
scenario\cite{guth1,guth2,rev1} has become an integral part of the standard
model of cosmology. However, to a great extent, this scenario is
incomplete. One of the problems is the lack of a model of inflation that is
adequate both from the inflationary {\em and} the particle physics viewpoints.
While many models\cite{models} exist that do all the things an inflationary
model needs to do, such as inflating, ending inflation gracefully, reheating
the universe and generating safe density fluctuations, none of these models
really is part of any reasonable extension of the standard model of particle
physics.

What we address in this work, however, is a more serious problem.  This has to
do with the dynamics of inflation, and more generally, the dynamics of scalar
fields in an expanding universe. By and large, the
various models of inflation make the assumption that the dynamics of the
spatial zero mode of the (so-called) inflaton field is governed by some
approximation to the effective potential which incorporates the effects of
quantum fluctuations of the field. Thus the equation of motion is usually of
the form:

\begin{equation}
\ddot{\phi} + 3\frac{\dot{a}}{a}\dot{\phi} + V'_{\rm eff}(\phi) = 0
\end{equation}

The problem here is that the effective potential is really only suited for
analyzing {\em static} situations; it is the effective action evaluated for a
field configuration that is constant in time \cite{effecpot}. Thus, it
is inconsistent to use the effective potential in a {\em dynamical}
situation. Notice that such inconsistency appears for {\bf any}
inflationary scenario (old, new, chaotic, ...).

More generally, the standard methods of high temperature field theory are
based on an equilibrium formalism\cite{dolan,sweinberg}; there is no time
evolution in such a
situation. Such techniques preclude us from treating non-equilibrium
situations
such as surely exist for very weakly coupled theories in the early universe.

In this work we try to rectify this situation by addressing three issues:
a) obtaining the evolution equations for the order parameter including
the {\it quantum fluctuations},
b) studying departures from thermal equilibrium if
the initial state is specified as a thermal ensemble, c) understanding the
renormalization aspects and the validity of the high temperature expansion.

Our ultimate goal is to study the dynamics of phase transitions in the early
universe, in particular, the formation and evolution of correlated domains and
symmetry breaking in an expanding universe.  From some of our previous studies
on the dynamics of phase transitions\cite{boy1,boy2} in Minkowski space, we
have learned that the familiar picture of ``rolling'' is drastically modified
when the fluctuations are taken into account. As the phase transition proceeds
fluctuations become large and correlated regions (domains) begin to grow. This
enhancement of the fluctuations modifies substantially the evolution equation
of the order parameter. Thus the time dependence of the order parameter is not
enough to understand the dynamical aspects of the phase transition; it must
be studied in conjunction with that of the fluctuations.  Vilenkin and
Ford\cite{vilford} and Linde\cite{linde1} studied the fluctuations in a free
scalar field theory in de Sitter space and Guth and Pi\cite{guthpi} studied the
growth of fluctuations by approximating a broken symmetry situation with an
inverted parabolic potential in de Sitter space.  However, we are not aware of
any previous attempt to incorporate the growth of fluctuations (arising from
the non-linearities) in the dynamics of the order parameter during cosmological
phase transitions.

Our approach is to use the functional Schr\"{o}dinger formulation, wherein we
specify the initial wavefunctional $\Psi[\Phi(\vec{.}); t]$
(or more generally a
density matrix $\rho[\Phi(\vec{.}), \tilde{\Phi}(\vec{.}); t]$), and then use
the Schr\"{o}dinger equation to evolve this state in time. We can then use
this state to compute all of the expectation values required in the
construction of the effective equations of motion for the order parameter of
the theory, as well as that for the fluctuations. The Schr\"{o}dinger approach
has already been used in the literature at zero temperature\cite{hill} and to
study non-equilibrium aspects of field theories\cite{eboli}.

One advantage of this approach is that it is truly a dynamical one; we set up
initial conditions at some time $t_o$ by specifying the initial state and then
we follow the evolution of the order parameter $\phi(t) \equiv
\langle\Phi(\vec{x})\rangle$ and of the fluctuations as this state evolves in
time. Another advantage is that it allows for departures from equilibrium.
Thus, issues concerning the restoration of symmetries in the early universe can
be addressed in a much more general setting.

There have been several attempts\cite{ringwald,leutwyler,weiss}  to obtain the
evolution equations in expanding cosmologies. Our motivations, goals and many
technical aspects differ substantially from those of previous treatments. In
particular, we not only obtain the evolution equations for the order parameter
to one-loop approximation, but we also find them in a non-perturbative
self-consistent Hartree approximation. Within these approximation schemes, we
obtain the evolution of the fluctuations (quantum and thermal),
 departures from equilibrium,  study
in detail the subtle  aspects of renormalization and the validity of a
high-temperature expansion.
Our analysis applies quite generally to any arbitrary spatially flat FRW
cosmology.
We also determine the time evolution of
the initial
(Boltzmann) distribution functions, relate the time evolution to ``squeezed
states'' and perform a numerical integration in the case of free fields for de
Sitter and radiation dominated cosmologies. We expect to provide a numerical
analysis of the evolution of the order parameter and the dynamics
of  phase transitions for interacting fields in a forthcoming article.

In the next section, we set up the formalism for
determining the dynamics of the
order parameter $\phi(t)$. This involves constructing the order $\hbar$
equations of motion for $\phi(t)$ incorporating quantum fluctuations, and then
constructing an {\em ansatz} for the time evolved density matrix we need to use
to evaluate the various expectation values in the problem. We then consider a
self-consistent (Hartree) approximation to the equations of motion (sec.3) and
deal with the issue of renormalization of these equations (sec.4).

The initial state we pick for the field $\Phi(\vec{x},t)$ is that corresponding
to a thermal density matrix centered at $\phi(t)$. It is then useful to try to
understand the high temperature limit of our calculations. We are able to
compute both the leading and subleading terms in the high $T$ expansion of
$\langle \phi^2(t)\rangle$. From this we show that the high $T$ expansion
cannot
be valid for all time, but breaks down in the large time limit (section 5).

In section 6 we compute the time evolution of the Boltzmann distribution
function (initially specified as thermal) as a function of time, and find
that to one-loop order and in the Hartree approximation, the density matrix
describes a ``squeezed'' state.  Section 7 provides a numerical analysis of the
departure from equilibrium in the simpler case of a free massive scalar field
in de Sitter and radiation dominated cosmologies. We point out that a
coarse-grained entropy used in the literature is {\em not} a monotonically
increasing function of time in the radiation dominated case. Section 8 contains
our conclusions. There are two appendices; the first contains some technical
results that are necessary in order compute the time evolved density matrix.
The second appendix treats some of the results of the paper in conformal
instead of comoving time.

\section{\bf Evolution Equations}

We start by setting up the {Schr\"{o}dinger} formalism for spatially flat FRW
cosmologies.
Consider a scalar field in such a cosmology where the metric is:
\begin{equation}
ds^2 = dt^2-a^2(t)d\vec{x}^2
\end{equation}
 The action and Lagrangian density are given by
\begin{eqnarray}
S         & =  & \int d^4x {\cal{L}} \label{action} \\
{\cal{L}} & =  & a^3(t)\left[\frac{1}{2}\dot{\Phi}^2(\vec{x},t)-\frac{1}{2}
\frac{(\vec{\nabla}\Phi(\vec{x},t))^2}{a(t)^2}-V(\Phi(\vec{x},t))\right]
 \label{lagrangian} \\
V(\Phi)   & =  & \frac{1}{2}[m^2+ \xi {\cal{R}}] \Phi^2(\vec{x},t)+
\frac{\lambda}{4!}\Phi^4(\vec{x},t) \label{potential} \\
{\cal{R}}    & =  & 6\left(\frac{\ddot{a}}{a}+\frac{\dot{a}^2}{a^2}\right)
\end{eqnarray}
with ${\cal{R}}$ the Ricci scalar.
The canonical momentum conjugate to $\Phi$ is
\begin{equation}
\Pi(\vec{x},t) = a^3(t)\dot{\Phi}(\vec{x},t) \label{canonicalmomentum}
\end{equation}
and the Hamiltonian becomes
\begin{equation}
H(t) = \int d^3x \left\{ \frac{\Pi^2}{2a^3(t)}+
\frac{a(t)}{2}(\vec{\nabla}\Phi)^2+
a^3(t) V(\Phi) \right\} \label{hamiltonian}
\end{equation}
In the Schr\"{o}dinger representation (at an arbitrary fixed time
$t_o$), the canonical momentum is represented
as
\[ \Pi(\vec{x}) = -i\hbar \frac{\delta}{\delta \Phi(\vec{x})} \]
Wave functionals obey the time dependent functional Schr\"{o}dinger equation
\begin{equation}
i\hbar \frac{\partial \Psi[\Phi,t]}{\partial t} = H \Psi[\Phi,t]
 \label{schroedinger}
\end{equation}

Since we shall eventually consider a ``thermal ensemble'' it is convenient to
work with a functional density matrix $\hat{\rho}$ with matrix elements in the
Schr\"{o}dinger representation
$\rho[\Phi(\vec{.}), \tilde{\Phi}(\vec{.});t]$. We will
{\it assume} that the
density matrix obeys the functional Liouville equation
\begin{equation}
i\hbar \frac{\partial \hat{\rho}}{\partial t} = \left[H(t),\hat{\rho}\right]
\label{liouville}
\end{equation}
whose formal solution is
\[ \hat{\rho}(t) = U(t,t_o) \hat{\rho}(t_o) U^{-1}(t,t_o)\] where
$U(t,t_o)$ is the time evolution operator, and $\hat{\rho}(t_o)$ the
density matrix at the arbitrary initial time $t_o$.

The diagonal density matrix elements $\rho[\Phi,\Phi;t]$ are interpreted
as a probability density in functional space.
Since we are considering an homogeneous and isotropic background, the
functional density matrix may be assumed to be translationally invariant.
 Normalizing the density matrix such that $Tr\hat{\rho}=1$, the
``order parameter'' is defined as
\begin{equation}
\phi(t) = \frac{1}{\Omega}\int d^3x \langle \Phi(\vec{x},t) \rangle =
\frac{1}{\Omega}\int d^3x  Tr\hat{\rho}(t)\Phi(\vec{x}) =
\frac{1}{\Omega}\int d^3x Tr \hat{\rho}(t_o)U^{-1}(t,t_o)\Phi(\vec{x},t_o)
U(t,t_o)
\label{orderparameter}
\end{equation}
where $\Omega$ is the comoving volume, and the scale factors cancel
between the numerator (in the integral) and the denominator. Note that we have
used the fact that the field operator does not evolve in time in this picture.
The evolution equations for the order parameter are the following
\begin{eqnarray}
\frac{d \phi(t)}{dt} & = & \frac{1}{a^3(t)\Omega}\int d^3x
  \langle \Pi(\vec{x},t) \rangle
 =\frac{1}{a^3(t)\Omega}\int d^3x  Tr\hat{\rho}(t)\Pi(\vec{x}) = \frac{\pi(t)}
{a^3(t)} \label{fidot} \\
\frac{d \pi(t)}{dt}     & = & -\frac{1}{\Omega}\int d^3x a^3(t) \langle
 \frac{\delta V(\Phi)}{\delta \Phi(\vec{x})} \rangle \label{pidot}
\end{eqnarray}
It is now convenient to write the field in the {Schr\"{o}dinger} picture as
\begin{eqnarray}
\Phi(\vec{x})   & = & \phi(t)+\eta(\vec{x},t) \label{split} \\
\langle \eta(\vec{x},t) \rangle
     & = & 0 \label{doteta}
\end{eqnarray}

Expanding the right hand side of (\ref{pidot}) we find the effective
equation of motion for the order parameter:
\begin{equation}
\frac{d^2 \phi(t)}{dt^2}+3 \frac{\dot{a}(t)}{a(t)}
\frac{d \phi(t)}{dt}+V'(\phi(t))+\frac{V'''(\phi(t))}{2 \Omega}\int d^3x
\langle \eta^2(\vec{x},t)\rangle+\cdots
=0 \label{effequation}
\end{equation}
where primes stand for derivatives with respect to $\phi$.
To leading order in the loop expansion we need that
$\langle \eta^2(\vec{x},t)\rangle = {\cal{O}}(\hbar)$. This will be guaranteed
to this order if the density matrix is assumed to be Gaussian with
a covariance (width) ${\cal{O}}(1/\hbar)$.
If (\ref{split}) is introduced in the Hamiltonian, we arrive at:
\begin{equation}
H(t) = \int d^3x \left\{-\frac{\hbar^2}{2 a^3(t)}\frac{\delta^2}{\delta
\eta^2}+\frac{a(t)}{2}\left(\vec{\nabla}\eta\right)^2+a^3(t)
\left(V(\phi)+V'(\phi)\eta+\frac{1}{2}V''(\phi)\eta^2+\cdots \right)
\right\} \label{quadham}
\end{equation}

Keeping only the terms quadratic in $\eta$ in (\ref{quadham})
gives the first order term in the loop expansion.

It is convenient to introduce the discrete Fourier transform of the
fields in the comoving frame as
\begin{equation}
\eta(\vec{x},t) = \frac{1}{\sqrt{\Omega}}\sum_{\vec{k}} \eta_{\vec{k}}(t)
e^{-i\vec{k}\cdot\vec{x}}
\label{fourier1}
\end{equation}
In this representation, the quadratic approximation to the Hamiltonian
(\ref{quadham}) becomes the Hamiltonian of a collection of independent harmonic
oscillators for each mode $\vec{k}$
\begin{eqnarray}
H_q & = & \Omega a^3(t) V(\phi(t))+\nonumber \\
 &   & \frac{1}{2} \sum_{\vec{k}}
\left\{ -\frac{\hbar^2}{a^3(t)}
 \frac{\delta^2}{\delta \eta_{\vec{k}}\delta\eta_{-\vec{k}}}+2a^3(t)
 V'_{\vec{k}}(\phi(t))\eta_{-\vec{k}}+
\omega^2_k(t)\eta_{\vec{k}}\eta_{-\vec{k}}\right\} \label{hamodes} \\
V'_{\vec{k}}(\phi(t))
 & = & V'(\phi(t))\sqrt{\Omega}\delta_{\vec{k},0} \nonumber \\
\omega^2_k(t)
 & = & a(t)\vec{k}^2+a^3(t)V''(\phi(t)) \label{timedepfreq}
\end{eqnarray}

We propose the following Gaussian ansatz for the functional density
matrix elements in the {Schr\"{o}dinger} representation
\begin{eqnarray}
\rho[\Phi,\tilde{\Phi},t] & = & \prod_{\vec{k}} {\cal{N}}_k(t) \exp\left\{
- \left[\frac{A_k(t)}{2\hbar}\eta_k(t)\eta_{-k}(t)+
\frac{A^*_k(t)}{2\hbar}\tilde{\eta}_k(t)\tilde{\eta}_{-k}(t)+
\frac{B_k(t)}{\hbar}\eta_k(t)\tilde{\eta}_{-k}(t)\right] \right. \nonumber \\
       &   & \left. +\frac{i}{\hbar}\pi_k(t)\left(\eta_{-k}(t)-
\tilde{\eta}_{-k}(t)\right) \right\} \label{densitymatrix} \\
\eta_k(t)          & = & \Phi_k-\phi(t)\sqrt{\Omega}\delta_{\vec{k},0}
\label{etaofkt} \\
\tilde{\eta}_k(t)       & = &
{\tilde{\Phi}}_k-\phi(t)\sqrt{\Omega}\delta_{\vec{k},0}
\label{etaprimeofkt}
\end{eqnarray}
where $\phi(t) = \langle \Phi(\vec{x}) \rangle$ and $\pi_k(t)$ is the Fourier
transform of $\langle \Pi(\vec{x}) \rangle$. This form of the density matrix
is dictated by the hermiticity condition $\rho^{\dagger}[\Phi,\tilde{\Phi},t] =
\rho^*[\tilde{\Phi},\Phi,t]$; as a result of this, $B_k(t)$ is real.
The kernel $B_k(t)$ determines the amount of ``mixing'' in the
density matrix, since if $B_k=0$, the density matrix corresponds to a pure
state because it is a wave functional times its complex conjugate.

In order to solve for the time evolution of the density matrix
(\ref{liouville}) we need to specify the density matrix at some initial
time $t_o$. It is at this point that we have to {\it assume} some physically
motivated initial condition. We believe that this is a subtle point that
has not received proper consideration in the literature. A system in
thermal equilibrium has time-independent ensemble averages (as the evolution
Hamiltonian commutes with the density matrix) and there is no memory of any
initial state. However, in a time dependent background, the density matrix
will evolve in time, departing from the equilibrium state and
correlation functions or expectation values may depend on
details of the initial state.

We will {\it assume} that at early times
the initial density matrix is {\em thermal} for the modes that
diagonalize the
Hamiltonian at $t_o$ (we call these the {\em adiabatic} modes). The effective
temperature for these modes is $k_BT_o = 1/\beta_o$. It is only in this
initial state that the notion of ``temperature'' is meaningful. As the
system departs from equilibrium one cannot define a thermodynamic temperature.
Thus in this case the ``temperature'' refers to the temperature defined in the
initial state.

The initial values of the
order parameter and average canonical momentum are $\phi(t_o) =\phi_o$ and
$\pi(t_o)=\pi_o$ respectively. Defining the adiabatic frequencies as:
\begin{equation}
 W^2_k(t_o) = \frac{\omega^2_k (t_o)}{a^3(t_o)}=
\frac{\vec{k}^2}{a^2(t_o)}+V''(\phi_{cl}(t_o)) \label{adiabfreq}
\end{equation}
we find that the initial values of the time dependent
parameters in the density matrix (\ref{densitymatrix}) are
\begin{eqnarray}
A_k(t_o) & = & A^*_k(t_o) =
 W_k(t_o)a^3(t_o)
\coth\left[\beta_o\hbar W_k(t_o)
\right] \label{Ato} \\
B_k(t_o) & = & - \frac{W_k(t_o)a^3(t_o)}
{\sinh\left[\beta_o\hbar W_k(t_o)\right]}
\label{Bto} \\
{\cal{N}}_k(t_o)
   & = &  \left[\frac{W_k(t_o)a^3(t_o)}{\pi\hbar}\tanh\left[
\frac{\beta_o\hbar W_k(t_o)}{2}\right]\right]^{\frac{1}{2}}
 \label{normo} \\
\phi(t_o)& = & \phi_o \; \; ; \; \; \pi(t_o)=\pi_o
\label{initialfielmom}
\end{eqnarray}
The initial density matrix is normalized such that
$Tr\rho(t_o)=1$. Since time evolution is unitary such a normalization
will be constant in time. For $T_o = 0$ the density matrix describes a
pure state since $B_k = 0$.

As an example, consider the case of de Sitter space. The scale factor is
given by $a(t) =a_o e^{Ht}$ and for $T_o \rightarrow 0$, $t_o \rightarrow
-\infty$ we recognize the
ground state wave-functional for the Bunch-Davies vacuum\cite{hill,birrell}.
For $T_o \neq 0$ this
initial density matrix corresponds to a thermal ensemble of
Bunch-Davies modes.
Certainly this choice is somewhat arbitrary but it physically describes the
situation in which at very early times the adiabatic modes are in {\it local
thermodynamic equilibrium}. Whether or not this situation actually obtains for
a given system has to be checked explicitly. In the cosmological setting,
the nature of the initial condition will necesarily have to result from a
deeper understanding of the relationship between particle physics, gravitation
and statistical mechanics at very large energy scales.

Although we will continue henceforth to use this thermal initial state, it
should be emphasized that our formalism is quite general and can be applied to
{\em any initial} state.

In the {Schr\"{o}dinger} picture, the Liouville equation (\ref{liouville})
becomes
\begin{eqnarray}
& & i\hbar \frac{\partial \rho[\Phi,\tilde{\Phi},t]}{\partial t} =
\sum_k\left\{
-\frac{\hbar^2}{2a^3(t)}\left(
\frac{\delta^2}{\delta \eta_k \delta \eta_{-k}}-
\frac{\delta^2}{\delta \tilde{\eta}_k \delta \tilde{\eta}_{-k}}\right) \right.
 \nonumber \\
& & \left. +a^3(t){V'}_{-k}(\phi(t))\left(\eta_{k}-\tilde{\eta}_k\right)+
\frac{1}{2}\omega^2_k(t)\left(\eta_{\vec{k}}\eta_{-\vec{k}}
-\tilde{\eta}_k\tilde{\eta}_{-k}\right)
\right\} \rho[\Phi,\tilde{\Phi},t] \label{liouvischroed}
\end{eqnarray}
Since the modes do not mix in this approximation to the Hamiltonian, the
equations for the kernels in the density matrix are obtained by comparing the
powers of $\eta$ on both sides of the above equation. We obtain the following
equations for the coefficients:
\begin{eqnarray}
i\frac{\dot{{\cal{N}}}_k}{{\cal{N}}_k} & = & \frac{1}{2a^3(t)}(A_k-A^*_k)
\label{normeq} \\
i\dot{A}_k                     & = &
\left[ \frac{A^2_k-B^2_k}{a(t)^3}-\omega^2_k(t)\right] \label{Aeq} \\
i \dot{B_k}                    & = & \frac{B_k}{a^3(t)}(A_k-A^*_k)
\label{Beq} \\
-\dot{\pi}_k                   & = & V'(\phi(t))a^3(t) \sqrt{\Omega}
\delta_{\vec{k},0} \label{pieq} \\
\dot{\phi}                     & = & \frac{\pi}{a^3(t)}
\label{fieq}
\end{eqnarray}
The last two equations are identified with the {\it classical}
 equations of motion for the order parameter (\ref{effequation}).
The equation for $B_k(t)$ reflects the fact that a pure state
$B_k=0$ remains pure under time evolution.

Writing $A_k$ in terms of its real and imaginary components
$A_k(t) = A_{Rk}(t)+i A_{Ik}(t)$ (and because $B_k$ is real) we
find that
\begin{equation}
\frac{B_k(t)}{A_{Rk}(t)} = \frac{B_k(t_o)}{A_{Rk}(t_o)}
\label{invar}
\end{equation}
and that  the time evolution is unitary (as it should be), that is
\begin{equation}
\frac{{\cal{N}}_k(t)}{\sqrt{\left(A_{Rk}(t)+B_k(t)\right)}} = \mbox{constant}
\label{unitarity}
\end{equation}

The initial conditions (\ref{Ato},\ref{Bto}) and the
invariance of the ratio (\ref{invar}) suggest that
the solution for the real part of $A$ and for $B$ may be obtained by
introducing a complex function ${\cal{A}}_k(t)={\cal{A}}_{Rk}(t)+i
{\cal{A}}_{Ik}(t)$
\begin{eqnarray}
{\cal{A}}_{Rk}(t)           & = & A_{Rk}(t)
\tanh\left[\beta_o\hbar W_k(t_o)\right] =
- B_k(t)\sinh\left[\beta_o\hbar W_k(t_o)\right]  \label{Aoftime} \\
{\cal{A}}_{Rk}(t_o)     & = & W_k(t_o)~a^3(t_o) \label{ARinitial} \\
{\cal{A}}_{Ik}(t)               & = & A_{Ik}(t) \label{imagpart}
\end{eqnarray}
In this form, the real and imaginary parts of ${\cal{A}}$ satisfy the
equations
\begin{eqnarray}
\dot{{\cal{A}}}_{Rk}(t)  & = & \frac{2}{a^3(t)}{\cal{A}}_{Rk}(t)
{\cal{A}}_{Ik}(t) \label{realeq} \\
-\dot{{\cal{A}}}_{Ik}(t) & = & \frac{1}{a^3(t)}\left[{\cal{A}}^2_{Rk}(t)-
{\cal{A}}^2_{Ik}(t) -\omega^2_k(t)a^3(t)\right] \label{imageq}
\end{eqnarray}
These two equations may be combined in one complex equation for the
combination
 ${\cal{A}}_k(t) = {\cal{A}}_{Rk}(t)+
i{\cal{A}}_{Ik}(t)$ that obeys the Riccati-type equation
\begin{equation}
i\dot{\cal{A}}_k(t) = \frac{1}{a^3(t)}\left[{{\cal{A}}_k}^2(t)-
\omega^2_k(t)a^3(t) \right] \label{riccati}
\end{equation}
with the initial conditions:
\begin{eqnarray}
{\cal{A}}_{Rk}(t_o) & = &  W_k(t_o)a^3(t_o) \label{Are} \\
{\cal{A}}_{Ik}(t_o) & = & 0 \label{Aim}
\end{eqnarray}

The Riccati equation (\ref{riccati}) becomes a more amenable differential
equation by the change of variables

\begin{equation}
{\cal{A}}_k(t) = -ia^3(t)\frac{\dot{\varphi}_k(t)}{\varphi_k(t)}
\label{changeofvar}
\end{equation}

The solution to the Ricatti equation with the above initial conditions
is detailed in  appendix A.
We find that it is convenient to introduce two real mode functions
(for each wavevector $k$) and write

\begin{equation}
\varphi_k(t) =
\frac{{\cal{U}}_{k1}(t)+i{\cal{U}}_{k2}(t)}{\sqrt{a^3(t)W_k(t_o)}}.
 \label{chanvar}
\end{equation}
These mode functions obey a {Schr\"{o}dinger}-like equation. The initial
conditions on $\varphi_k(t_o)$ are (see appendix A)

\begin{eqnarray}
\varphi_k(t_o)               & = & \frac{1}{\sqrt{a^3(t_o)W_k(t_o)}}
 \nonumber \\
\dot{\varphi}_k(t)\mid_{t_o} & = &
 i\sqrt{\frac{W_k(t_o)}{a^3(t_o)}}
\nonumber
\end{eqnarray}

 ${\cal{A}}_{Rk}(t)$ is given by equation
 (\ref{arealfin}) in  appendix A, so that:

\begin{eqnarray}
A_{Rk}(t) & = &
 \frac{1}{\mid \varphi_k(t) \mid^2}
\coth\left[\beta_o\hbar W_k(t_o)
\right] \label{Aoftimefin} \\
B_k(t)    & = & -
\frac{1}{\mid \varphi_k(t) \mid^2}
\left[\frac{1}{\sinh\left[\beta_o\hbar W_k(t_o)\right]}\right]
\label{Boftimefin}
\end{eqnarray}

The equal time two-point function for the fluctuation becomes
\begin{equation}
\langle \eta_{k}(t)\eta_{-k}(t) \rangle = \frac{\hbar}{2\left(
A_{Rk}(t)+B_{k}(t)\right)} =
\frac{\hbar}{2}{\mid \varphi_k(t) \mid^2}
\coth\left[\beta_o\hbar W_k(t_o)/2
\right] \label{twopointfunc}
\end{equation}
The one-loop equation of motion for the order parameter thus becomes

\begin{equation}
\ddot{\phi}+3 \frac{\dot{a}}{a}\dot{\phi}+V'(\phi)+V'''(\phi)
\frac{\hbar}{2}\int \frac{d^3k}{(2\pi)^3}
\frac{\mid \varphi_k(t) \mid^2}{2}
\coth\left[\beta_o\hbar W_k(t_o)/2
\right] = 0 \label{finaleqofmot}
\end{equation}
with the function $\varphi_k(t)$ defined  in  appendix A by   (\ref{calU},
\ref{diffeqU}) in which, to this order in $\hbar$, only the {\it classical
solution}  $\phi_{cl}(t)$ enters. A consistent numerical solution of these
equations to ${\cal{O}}(\hbar)$ would involve splitting
$\phi(t)=\phi_{cl}(t)+\hbar \phi_{(1)}(t)$ and keeping only the
${\cal{O}}(\hbar)$ terms in the evolution equation. This will result in two
simultaneous equations, one for the classical evolution of the order
parameter and another for $\phi_{(1)}(t)$.

This equation of motion is clearly {\em very} different from the one obtained
by using the effective potential. It may be easily seen (by writing the
effective action as the classical action plus the logarithm of the determinant
of the quadratic fluctuation operator) that this is the equation of motion
obtained by the variation of the one-loop effective action.

The {\it static} effective potential is clearly not the appropriate quantity to
use to describe scalar field dynamics in an expanding universe. Although there
may be some time regime in which the time evolution is slow and fluctuations
rather small, this will certainly {\it not} be the case at the onset of a phase
transition. As the phase transition takes place, fluctuations become dominant
and grow in time signaling the onset of long range
correlations\cite{boy1,boy2}.

\section{\bf Hartree equations}

Motivated by our previous studies in Minkowski space\cite{boy1,boy2} which
showed that the growth of correlation and enhancement of fluctuations during
a phase transition may not be described perturbatively, we now proceed to
obtaining the equations of motion in a Hartree approximation. This
approximation is non-perturbative in the sense that it sums up infinitely
many diagrams of the cactus-type\cite{dolan}. The Hartree approximation
becomes exact in the $N \rightarrow \infty$ limit of an $O(N)$ vector theory.
Although its validity is not warranted in the present case, it at least
provides a consistent non-perturbative framework in which correlations
and fluctuations can be studied. It is conceivable that this
approximation could be implemented beyond the lowest
(cactus) order in a consistent fashion.

Vilenkin\cite{vilenkin} has previously studied a simplified version of the
Hartree approximation in which, however, the mode functions that enter in the
propagators did not incorporate the fluctuations in a  self-consistent manner.

The Hartree self-consistent approximation is implemented as follows. We
decompose the field as in (\ref{split}), using a potential as in
(\ref{potential}). We find that the Hamiltonian becomes
\begin{eqnarray}
H & = &  \int d^3x \left\{-\frac{\hbar^2}{2 a^3(t)}\frac{\delta^2}
{\delta\eta^2}+\frac{a(t)}{2}\left(\vec{\nabla}\eta\right)^2+a^3(t)
\left(V(\phi)+V'(\phi)\eta+\frac{1}{2!}V''(\phi)\eta^2 \right. \right.
\nonumber \\
  & + &
\left. \left. \frac{1}{3!} \lambda \phi\eta^3+
\frac{1}{4!} \lambda \eta^4 \right)\right\} \label{fullham}
\end{eqnarray}

The Hartree approximation is obtained by assuming the factorization
\begin{eqnarray}
\eta^3 (\vec{x},t) &  \rightarrow & 3 \langle \eta^2(\vec{x},t) \rangle
\eta(\vec{x},t) \label{eta3} \\
\eta^4 (\vec{x},t) &  \rightarrow & 6 \langle \eta^2(\vec{x},t) \rangle
\eta^2(\vec{x},t) -3
\langle \eta^2(\vec{x},t) \rangle^2 \label{eta4}
\end{eqnarray}
where $\langle \cdots \rangle$ is the average using the time evolved density
matrix. This average will be determined self-consistently (see below).
Translational invariance shows that $\langle \eta^2(\vec{x},t) \rangle$ can
only be a function of time. This approximation makes the Hamiltonian quadratic
at the expense of a self-consistent condition. In the time independent
(Minkowski) case this approximation sums up all the ``daisy'' (or ``cactus'')
diagrams and leads to the self-consistent gap equation\cite{dolan}. In this
approximation the Hamiltonian becomes
\begin{eqnarray}
& & H = \Omega a^3(t) {\cal{V}}(\phi) +  \nonumber \\
& &  \int d^3x \left\{-\frac{\hbar^2}{2 a^3(t)}\frac{\delta^2}{\delta
\eta^2}+ \frac{a(t)}{2}\left(\vec{\nabla}\eta\right)^2
+a^3(t)\left({\cal{V}}^{(1)}(\phi)\eta+\frac{1}{2}
{\cal{V}}^{(2)}(\phi)\eta^2 \right)
\right\} \label{hartham}\\
& & {\cal{V}}(\phi) = V(\phi)- \frac{1}{8}\lambda \langle \eta^2 \rangle^2
\label{newV} \\
& & {\cal{V}}^{(1)}(\phi) = V'(\phi)+\frac{1}{2}\lambda\phi
\langle \eta^2 \rangle \label{newVprime} \\
& & {\cal{V}}^{(2)}(\phi) = V''(\phi)+ \frac{1}{2}
\lambda  \langle \eta^2 \rangle \label{newVdoubprim}
\end{eqnarray}

We can now introduce the Fourier transform of the field as in
(\ref{fourier1}).The Hamiltonian will have the same form as (\ref{hamodes}) but
the time dependent frequencies (\ref{timedepfreq}) and linear term in $\eta$
become
\begin{eqnarray}
\omega_k^2(t) & = & a(t) \vec{k}^2+a^3(t){\cal{V}}^{(2)}(\phi(t))
 \label{newfreq} \\
{\cal{V}}^{(1)}_{\vec{k}}(\phi(t))
     & = & {\cal{V}}^{(1)}(\phi(t))
\sqrt{\Omega}\delta_{\vec{k},0}
\end{eqnarray}
The ansatz for the Gaussian density matrix is the same as before
(\ref{densitymatrix}), as are the evolution equations for the coefficients
$A_k(t), \ B_k(t), \ {\cal{N}}_k(t)$.
However, the frequencies are now given by (\ref{newfreq}). The {\it classical}
equations of motion (\ref{pieq},\ref{fieq}) now become
\begin{eqnarray}
-\dot{\pi}                     & = & {\cal{V}}^{(1)}(\phi(t))a^3(t)
 \label{newpieq} \\
\dot{\phi}                     & = & \frac{\pi}{a^3(t)}
\label{newfieq}
\end{eqnarray}
The equations for the coefficients $A_k(t), \  B_k(t), \ {\cal{N}}_k(t)$ are
again solved in terms of the mode functions given in  appendix A but with
$V''(\phi_{cl}(t))$ now replaced by ${\cal{V}}^{(2)}(\phi(t))$ and the
following
replacement of the adiabatic frequencies:

\begin{eqnarray}
 W_k(t_o) \rightarrow {\cal{W}}_k(t_o) & = &
 {{\left[ {\vec{k}^2}+
{m^2(T_o)} \right]^{\frac{1}{2}}}\over {a(t_o)}}
\label{hartfreq} \\
\frac{m^2(T_o)}{a^2(t_o)}              & = & {\cal{V}}^{(2)}(\phi(t_o))
\label{defmass}
\end{eqnarray}

The mode functions of the appendix obey the differential equations and initial
conditions with this replacement and \[\mid \varphi_k(t) \mid^2 \rightarrow
\mid \varphi^H_k(t) \mid^2 .\] The
initial conditions at $t_o$ are now given in terms of these new adiabatic
frequencies (see appendix A).  The equal time two-point function thus becomes
\begin{equation}
\langle \eta^2(\vec{x},t) \rangle =  \frac{\hbar}{2}
 \int \frac{d^3k}{(2\pi)^3}
\mid \varphi^H_k(t) \mid^2
\coth\left[\beta_o\hbar {\cal{W}}_k(t_o)/2
\right], \label{newtwopointfunc}
\end{equation}
which leads to the following set of self-consistent time dependent
Hartree equations:
\begin{eqnarray}
& & \ddot{\phi}+3 \frac{\dot{a}}{a}\dot{\phi}+V'(\phi)+
\lambda \phi \frac{\hbar}{2}  \int \frac{d^3k}{(2\pi)^3}
\frac{\mid \varphi^H_k(t) \mid^2}{2}
\coth\left[\beta_o\hbar {\cal{W}}_k(t_o)/2
\right] = 0 \label{harteqofmot} \\
& & \left[\frac{d^2}{dt^2}+3 \frac{\dot{a}(t)}{a(t)}\frac{d}{dt}+
\frac{\vec{k}^2}{a^2(t)}+V''(\phi(t)) + \right. \\
& & \left. \lambda \frac{\hbar}{2}
 \int \frac{d^3k}{(2\pi)^3}
\frac{\mid \varphi^H_k(t) \mid^2}{2}
\coth\left[\beta_o\hbar {\cal{W}}_k(t_o)/2
\right] \right]
\varphi^H_k(t)
=0 \label{newdiffeqU} \\
& &
\varphi^H_k(t_o)        =  \frac{1}{\sqrt{a^3(t_o){\cal{W}}_k(t_o)}}
\label{newboundconU} \\
& &
\dot{\varphi}^H_k(t)\mid_{t_o}  =
 i\sqrt{\frac{{\cal{W}}_k(t_o)}{a^3(t_o)}}
\label{newboundconUdot}
\end{eqnarray}

\section{Renormalization}

In either the one-loop approximation or the Hartree self-consistent
approximation, the renormalization aspects are contained in the momentum
integrals of the mode functions. Because the Bose-Einstein distribution
functions are exponentially suppressed at large momenta, the finite
temperature contribution will be convergent and we need only address
the zero temperature contribution. The study of renormalization is more
conveniently performed in terms of the following mode function
satisfying a {Schr\"{o}dinger}-like equation (see Appendix A):
\begin{eqnarray}
{\cal{U}}^{H}_k(t)
         & = & a^{\frac{3}{2}}(t)\varphi^H_k(t)
\sqrt{{\cal{W}}_k(t_o)} \label{ucomplex} \\
{\cal{U}}^{H}_k(t_o)      & = & 1 \label{boundcomplex} \\
{\dot{\cal{U}}}^{H}_k(t)  & = & \frac{3\dot{a}(t_o)}{2a(t_o)}
+ i {\cal{W}}_k(t_o) \label{dotboundcomp}
\end{eqnarray}
The one-loop case may be obtained by making the replacement
${\cal{W}}_k(\phi)\rightarrow W_k(\phi)$.

We will now analyze the renormalization aspects for the Hartree
approximation; the one-loop case may be obtained easily from this more
general case.
We need to understand the divergences in the integral
\begin{equation}
I =  \int \frac{d^3k}{(2\pi)^3}
\frac{\mid {\cal{U}}^{H}_{k}(t) \mid^2}
{2a^3(t){\cal{W}}_k(t_o)} \label{divint}
\end{equation}
The divergences in this integral will be determined from the large-k
behavior of the mode function that is a solution to the differential
equation obtained from (\ref{newdiffeqU}) with the initial conditions
(\ref{boundcomplex}, \ref{dotboundcomp}). The large-k behavior of this
function may be obtained in a WKB approximation by introducing the
function ${\cal{D}}_{ k}(t)$,
\begin{eqnarray}
{\cal{D}}_k(t)   & = & \exp\left[\int^t_{t_o} R(t')dt'\right] \label{wkb} \\
{\cal{D}}_k(t_o) & = & 1 \label{wkbboundcon}
\end{eqnarray}
satisfying the differential equation
\begin{equation}
\left[\frac{d^2}{dt^2}-\frac{3}{2}\left(\frac{\ddot{a}}{a}+
\frac{1}{2}
\frac{\dot{a}^2}{a^2}\right)+
\frac{\vec{k}^2}{a^2(t)}+V''(\phi(t)) +
\frac{\lambda}{2} \langle \eta^2(\vec{x},t) \rangle \right]
{\cal{D}}_{ k}(t)
=0 \label{wkbequ}
\end{equation}
with $ \langle \eta^2(\vec{x},t) \rangle $ being the self-consistent integral
in the Hartree equation (\ref{newdiffeqU}). The one-loop approximation is
obtained by setting $\langle \eta^2(\vec{x},t) \rangle =0$ in the above
equation. The mode function ${\cal{U}}^{H}$ is obtained as a linear combination
of the function ${\cal{D}}_k(t)$  and its complex conjugate; the
coefficients are to be determined from the initial conditions (see below). The
function $R(t)$ obeys a Riccati  equation
\begin{equation}
\dot{R}+R^2-\frac{3}{2}\left(\frac{\ddot{a}}{a}+
\frac{1}{2}
\frac{\dot{a}^2}{a^2}\right)+
\frac{\vec{k}^2}{a^2(t)}+V''(\phi(t)) +
\frac{\lambda}{2} \langle \eta^2(\vec{x},t) \rangle = 0
\label{wkbriccati}
\end{equation}
We propose a WKB solution to this equation of the form
\begin{equation}
R= \frac{-i k}{a(t)}+R_o(t)-\frac{i R_1(t)}{k}+\frac{R_2(t)}{k^2}+\cdots
\label{wkbseries}
\end{equation}
and find the time dependent coefficients by comparing powers of $k$.
 This yields:
\begin{eqnarray}
R_o(t) & = & \frac{\dot{a}(t)}{2a(t)} \label{Ro} \\
R_1(t) & = & \frac{a(t)}{2}\left[-\frac{{\cal{R}}}{6}+V''(\phi(t))+
\frac{\lambda}{2} \langle \eta^2(\vec{x},t) \rangle \right] \label{R1} \\
R_2(t) & = & -\frac{1}{2}\frac{d}{dt}\left[a(t)R_1(t)\right] \label{R2}
\end{eqnarray}
Finally, we write
\begin{equation}
{\cal{U}}^{H}_k(t)  =  \frac{1}{2}[1+\gamma]{\cal{D}}^*_k(t)+
\frac{1}{2}[1-\gamma]{\cal{D}}_k(t) \label{combo}
\end{equation}
where $\gamma$ is determined from the initial condition
(\ref{dotboundcomp}):
\begin{equation}
\gamma = 1- i \frac{\dot{a}(t_o)}{k}+
\frac{ m(T_o)^2}{2 k^2}-\frac{a(t_o)
R_1(t_o)}{k^2} + {\cal{O}}(1/k^3)+\cdots
\label{coeff}
\end{equation}
 Thus in the $k \rightarrow \infty$ limit we find
\begin{equation}
\frac{\mid \varphi^H_k(t) \mid^2}{2} \rightarrow
 \left\{\frac{1}{2a^2(t)k}+\frac{1}{4k^3}
\left[\frac{\dot{a}^2(t_o)}{a^2(t)}-\left(-\frac{{\cal{R}}}{6}+V''(\phi)+
\frac{\lambda}{2} \langle \eta^2(\vec{x},t) \rangle \right)\right]+
 {\cal{O}}(1/k^4)+\cdots\right\} \label{divin}
\end{equation}

Introducing an upper momentum cut-off $\Lambda$ we obtain
\begin{eqnarray}
\langle \eta^2 (\vec{x},t) \rangle =
 &  & \hbar \int \frac{d^3k}{(2\pi)^3}
\frac{\mid \varphi^H_k(t) \mid^2}{2}
\coth\left[\beta_o\hbar {\cal{W}}_k(t_o)/2
\right] = \frac{\hbar}{8\pi^2} \frac{\Lambda^2}{a^2(t)}+ \nonumber \\
 &  & \frac{\hbar}{8\pi^2}
\ln \left(\frac{\Lambda}{K}\right)
 \left[\frac{\dot{a}^2(t_o)}{a^2(t)}-\left(-\frac{{\cal{R}}}{6}+
V''(\phi(t))+\frac{\lambda}{2} \langle \eta^2(\vec{x},t) \rangle
\right)\right]+
\mbox{ finite } \label{divergences}
\end{eqnarray}
where we have introduced a renormalization point $K$, and the finite part
depends on time, temperature and $K$. In the one-loop approximation, $\langle
\eta^2(\vec{x},t) \rangle $  does not appear in the logarithmic divergent term
in (\ref{divergences}) (as it does not appear in the differential equation for
the mode functions up to one loop).

There are several physically important features of the divergent structure
obtained above. First, the quadratically divergent term reflects the fact that
the physical momentum cut-off is being red-shifted by the expansion. This term
will not appear in dimensional regularization.

Secondly, the logarithmic divergence contains a term that reflects the initial
condition (the derivative of the expansion factor at the initial time $t_o$).
The initial condition breaks any remnant symmetry. For example, in de Sitter
space there is still invariance under the de Sitter group, but this is also
broken by the initial condition at an arbitrary time $t_o$.Thus this term is
not forbidden, and its appearance does not come as a surprise. As a consequence
of this term, we need a time dependent term in the bare mass proportional to
$1/{a^2(t)}$.

We are now in a position to present the renormalization prescription within
the Hartree approximation. In this approximation there are no interactions,
since
the Hamiltonian is quadratic. The non-linearities are encoded in the
self-consistency condition. Because of this, there are
no counterterms with which to cancel the divergences and the differential
equation for the mode functions (\ref{newdiffeqU})  {\it must be finite}.
Thus the renormalization conditions are obtained from
\begin{equation}
m^2_B(t) + \frac{\lambda_B}{2}\phi^2(t)+\xi_B{\cal{R}}+\frac{\lambda_B}{2}
\langle \eta^2 \rangle_B =
m^2_R+\frac{\lambda_R}{2}\phi^2(t)+\xi_R{\cal{R}}+\frac{\lambda_R}{2}
\langle \eta^2 \rangle_R \label{renormcond}
\end{equation}
where the subscripts $B,\ R$ refer to bare and
 renormalized quantities
respectively and $\langle \eta^2 \rangle_B$ is read from
 (\ref{divergences})
\begin{equation}
\langle \eta^2 \rangle_B =  \frac{\hbar}{8\pi^2} \frac{\Lambda^2}{a^2(t)}+
 \frac{\hbar}{8\pi^2}
\ln \left(\frac{\Lambda}{K}\right)
 \left[\frac{\dot{a}^2(t_o)}{a^2(t)}-\left(-\frac{{\cal{R}}}{6}+
m^2_R+\frac{\lambda_R}{2}\phi^2(t)+\xi_R{\cal{R}}+\frac{\lambda_R}{2}
\langle \eta^2 \rangle_R \right)\right]+ \mbox{ finite }
\label{eta2}
\end{equation}
Using the renormalization conditions (\ref{renormcond}) we obtain
\begin{eqnarray}
& & m^2_B(t) +\frac{\lambda_B\hbar}{16\pi^2}\frac{\Lambda^2}{a^2(t)}+
\frac{\lambda_B\hbar}{16\pi^2}\ln \left(\frac{\Lambda}{K}\right)
 \frac{\dot{a}^2(t_o)}{a^2(t)}  =
m^2_R\left[1+\frac{\lambda_B\hbar}{16\pi^2}
\ln \left(\frac{\Lambda}{K}\right)\right] \label{massren} \\
& & \lambda_B                       =
\frac{\lambda_R}{1-\frac{\lambda_R\hbar}{16\pi^2}
\ln \left(\frac{\Lambda}{K}\right)} \label{lambdaren} \\
& & \xi_B                         =
\xi_R + \frac{\lambda_B\hbar}{16\pi^2}
\ln \left(\frac{\Lambda}{K}\right) \left(\xi_R-\frac{1}{6}\right)
\label{xiren} \\
& & \langle \eta^2 \rangle_R        =   I_R + J
\label{etaren}
\end{eqnarray}
where
\begin{eqnarray}
I_R & = & \hbar\int \frac{d^3k}{(2\pi)^3} \left\{
\frac{\mid \varphi^H_k(t) \mid^2}{2} -
\frac{1}{2ka^2(t)} + \right. \nonumber \\
 &   & \left. \frac{\theta(k-K)}{4k^3}\left[
-\frac{{\cal{R}}}{6}- \frac{\dot{a}^2(t_o)}{a^2(t)}
+ m^2_R+\frac{\lambda_R}{2}\phi^2(t)+\xi_R{\cal{R}}+\frac{\lambda_R}{2}
\langle \eta^2 \rangle_R
\right] \right\}
\label{irren} \\
J   & = & \hbar\int \frac{d^3k}{(2\pi)^3}
\frac{\mid \varphi^H_k(t)\mid^2}{\exp{\beta_o\hbar {\cal{W}}_k(t_o)} - 1}
\label{jota}
\end{eqnarray}

The conformal coupling $\xi = 1 / 6$ is a {\it fixed point} under
renormalization\cite{birrell}. In dimensional regularization the terms
involving $\Lambda^2$ are absent and $\ln\Lambda$ is replaced by a simple pole
at the physical dimension. Even in such a regularization scheme, however, a
time dependent bare mass is needed. The presence of this new renormalization
allows us to introduce a new renormalized mass term  of the form \[
\frac{\Sigma}{a^2(t)} \] This counterterm may be interpreted as a squared mass
red-shifted by the expansion of the universe. However, we shall set $\Sigma =
0$ for simplicity.

It is clear that there is no wave function renormalization. This is a
consequence of the approximations invoked. There is, in fact, no wavefunction
renormalization in either the one-loop or the Hartree approximation for a
scalar field theory in three spatial dimensions.

Notice that there is a weak cut-off dependence on the effective equation
of motion for the order parameter.

 For {\it fixed} $\lambda_R$,
as the cutoff $\Lambda \rightarrow \infty$
\begin{eqnarray}
  \lambda_B \approx -\frac{(4\pi)^2}{\ln\left(\frac{\Lambda}{K}\right)}
\label{lamdab}
\end{eqnarray}
In addition,
\begin{eqnarray}
& & \xi_B = \frac{1}{6} + O\left(\frac{1}{\ln\Lambda}\right)
\label{xibare} \\
& & m^2_B(t) = \frac{1}{a^2(t)}\left[\frac{\Lambda^2}{\ln
\left(\frac{\Lambda}{K}\right)} +  {\dot{a}}^2(t_0)  \right]+
O\left(\frac{1}{\ln\Lambda}\right)
\label{mbaretime}
\end{eqnarray}

This approach to $0^-$ of the bare coupling as the cutoff is removed
translates into an instability in the bare theory. This is a consequence of the
fact that the N-component $\Phi^4$ theory for $N \to \infty$ is asymptotically
free (see ref.\cite{asympto}), which is not relieved in curved space-time.
Clearly this theory is sensible only as a low-energy cut-off effective theory,
and it is in this restricted sense that we will ignore the weak cut-off
dependence and neglect the term proportional to the bare coupling in
(\ref{renharteqofmot}).

The renormalized self-consistent Hartree equations thus become
after letting $\Lambda = \infty$:
\begin{eqnarray}
& & \ddot{\phi}+3 \frac{\dot{a}}{a}\dot{\phi}+m^2_R\phi+\xi_R{\cal{R}}
\phi+ \frac{\lambda_R}{2} \phi^3+
\frac{\lambda_R \phi}{2} \langle \eta^2 \rangle_R =0
 \label{renharteqofmot} \\
& & \left[\frac{d^2}{dt^2}+3 \frac{\dot{a}(t)}{a(t)}\frac{d}{dt}+
\frac{\vec{k}^2}{a^2(t)}+m^2_R+ \xi_R {\cal{R}}+ \frac{\lambda_R}{2}
\phi^2 +
 \frac{\lambda_R}{2} \langle \eta^2 \rangle_R \right]
 \varphi^H_k(t) =0 \label{rennewdiffeqU}
\end{eqnarray}
where $\langle \eta^2 \rangle_R$ is given by equations (\ref{irren},
\ref{jota}).

For completeness we quote the renormalized equation of motion for the
order  parameter {\it to one-loop}

\begin{eqnarray}
& & \ddot{\phi}+3 \frac{\dot{a}}{a}\dot{\phi}+m^2_R\phi+\xi_R{\cal{R}}
\phi+ \frac{\lambda_R}{2} \phi^3+
\frac{\lambda_R \phi}{2} \langle \eta^2 \rangle_R  = 0
  \label{1lupreneqofmot} \\
& & \left[\frac{d^2}{dt^2}+3\frac{\dot{a}(t)}{a(t)}\frac{d}{dt}+
\frac{\vec{k}^2}{a^2(t)}+m^2_R+ \xi_R {\cal{R}}+ \frac{\lambda_R}{2}
\phi^2(t)  \right]
 \varphi_k(t) =0 \label{1luprennewdiffeqU} \\
\end{eqnarray}

The one-loop coupling constant renormalization differs from that in the Hartree
approximation by a factor of three. This is a consequence of the fact that the
Hartree approximation is equivalent to a large N approximation and only sums up
the s-channel bubbles.

\section{High Temperature Limit}

One of the payoffs of understanding the large-$k$ behavior of the mode
functions (as obtained in the previous section via the WKB method) is
that it permits the evaluation of the high temperature limit. We shall
perform our analysis of the high temperature expansion for the Hartree
approximation. The one-loop case may be read off from these results.

The finite temperature contribution is determined by the integral

\begin{equation}
J =
\hbar
\int \frac{d^3k}{(2\pi)^3}
\frac{\mid \varphi^H_k(t) \mid^2}{e^{\beta_o
\hbar {\cal{W}}_k(t_o)}-1}  \label{tempint}
\end{equation}
For large temperature, only momenta $k \geq T_o$ contribute. Thus the
leading contribution is determined by the first term of the function
$R(t)$ (eq.(\ref{wkbseries})) of the previous section.
 We find
\begin{equation}
J = \frac{1}{12 \hbar}
\left[ \frac{k_B T_o a(t_o)}{a(t)}\right]^2
\left[1+{\cal{O}}(1/T_o) + \cdots \right] \label{highT}
\end{equation}

Thus we see that the leading high temperature behavior reflects the physical
red-shift in the cosmological background and it results in an effective time
dependent temperature

\[ T_{eff}(t) = T_o \left[\frac{a(t_o)}{a(t)}\right] \]

To leading order, the expression obtained for the time dependent effective
temperature corresponds to what would be obtained for an {\it adiabatic}
(isentropic) expansion for blackbody-type radiation consisting of massless
relativistic particles evolving in the cosmological background.

This behavior only appears at {\it leading} order in the high temperature
expansion. There are subleading terms that must be taken into account. These
can be calculated within the high temperature expansion and we do this below.
To avoid cluttering of notation, we will set $k_B=\hbar=1$ in what follows.

We define
\begin{equation}
m^2(T_o)\equiv m^2_R+ \xi_R {\cal{R}}(t_o)+ \frac{\lambda_R}{2}
\phi^2(t_o)  + \frac{\lambda_R}{2}\langle \eta^2(t_o) \rangle_R
\end{equation}
and we will assume that $m^2(T_o) \ll T_o^2$.  Since we are interested in
the description of a phase transition, we will write
\begin{equation}
m^2_R+ \xi_R {\cal{R}}(t_o)+ \frac{\lambda_R}{2}\phi^2(t_o)  =
-\frac{\lambda_R T^2_c}{24} ; \; \; \;  T^2_c >0 \label{Tc}
\end{equation}

Thus, to leading order in $T_o$
\begin{equation}
 m^2(T_o) = \frac{\lambda_R}{24}(T_o^2-T^2_c)  \label{mofT}
\end{equation}
Our high temperature expansion will assume {\it fixed} $m(T_o)$ and
$m(T_o)/T_o \ll 1$.

It becomes  convenient to define the variable
\begin{eqnarray}
& & x^2 = \frac{k^2}{T_o^2 a^2(t_o)}+\frac{m^2(T_o)}{T_o^2} \nonumber \\
& & \mid \varphi^H_k(t) \mid^2 = \left|  \varphi^H(a(t_o))
\sqrt{x^2T_o^2-m^2(T_o)};t)
\right |^2
 \nonumber
\end{eqnarray}
Recall from our WKB analysis that the leading behavior for
$k\rightarrow \infty$ is (see equation (\ref{divin})
\[ \frac{\mid \varphi^H_k(t) \mid^2}{2} \rightarrow \frac{1}{2a^2(t)k} \]
adding and subtracting this leading term in the integral $J$ and performing
the
above change of variables, we have

\begin{eqnarray}
   J & = & J_1+J_2 \nonumber \\
 J_1 & = & \left[\frac{a(t_o)}{a(t))}\right]^3
\left(\frac{T_o}{\pi}\right)^2 \int^{\infty}_{\frac{m(T_o)}{T_o}}
\frac{x dx}{e^x-1}
\left[a^3(t)\sqrt{x^2T_o^2-m^2(T_o)}\frac{\mid \varphi^H_k(t)
\mid^2}{2}-\frac{a(t)}{2a(t_o)}
\right] \label{J1} \\
 J_2 & = & \frac{T_o^2}{2\pi^2}
\left[\frac{a(t_o)}{a(t)}\right]^2
\int^{\infty}_{\frac{m(T_o)}{T_o}}
\frac{x dx}{e^x-1}  \nonumber \\
  & = &  \frac{1}{2}T_o^2 \left[\frac{a(t_o)}{a(t)}\right]^2
\left[\frac{1}{12}-\frac{m(T_o)}{2\pi^2T_o}+\frac{m^2(T_o)}{8\pi^2T_o^2}+
{\cal{O}}\left(\frac{m^3(T_o)}{T_o^3}\right)+\cdots \right] \label{J2}
\end{eqnarray}

We now must study the high temperature expansion of $J_1$. We will
restrict ourselves to the determination of the linear and logarithmic
dependence on $T_o$.
For this purpose, it becomes convenient to introduce yet another
change of variables
\[ x = \frac{m(T_o)}{T_o}z \]
and use the fact that in the limit $ T_o \gg m(T_o)$,
\[\frac{z}{e^{\frac{m(T_o)}{T_o}z}-1} \approx \frac{T_o}{m(T_o)}[1-
\frac{m(T_o)}{2T_o}z + \cdots ] \]
This yields the following
linear and logarithmic terms in $T_o$:
\begin{equation}
J_{1lin} = \left[\frac{a(t_o)}{a(t)}\right]^3 \frac{T_o m(T_o)}{\pi^2}
\int^{\infty}_{1} dz \left\{a^3(t) m(T_o)\sqrt{z^2-1}\frac{\mid \varphi^H_k(t)
\mid^2}{2}-
\frac{a(t)}{2a(t_o)} \right\}
\label{linearT}
\end{equation}
Note that the above integral is finite.

The logarithmic contribution is obtained by keeping the
${\cal{O}}(1/k^3)$ in the large momentum expansion of $\mid \varphi^H_k(t)
\mid^2$
given by equation (\ref{divin}) (in terms of the new variable $z$).
We obtain after some straightforward algebra:

\begin{equation}
J_{1log} = -\frac{\ln\left[\frac{m(T_o)}{T_o}\right]}{8\pi^2}
\left\{-\frac{{\cal{R}}}{6}-\frac{\dot{a}^2(t_o)}{a^2(t)}+
\left[\frac{a(t_o)}{a(t)}\right]^2\left[m^2(T_o)+
\frac{\lambda_R T^2_c}{24}\right] -\frac{\lambda_R T^2_c}{24}\right\}
\end{equation}
That is, in the limit $T_o >> m(T_o),~ J_1 = J_{1lin} + J_{1log} +
O((T_o)^0)$.

Comparing the ${\cal{O}}(T_o^2, T_o, \ln T_o)$ contributions it becomes
clear that they have very different time dependences through the
scale factor $a(t)$. Thus the high temperature expansion as presented
will not remain accurate at large times since the term quadratic in $T_o$
may become of the same order or smaller than the linear or logarithmic terms.
The high
temperature expansion and the long time limit are thus not interchangeable,
and any high temperature expansion is thus bound to be valid only within
some time regime that depends on the initial value of the temperature and
the
initial conditions.

As an illustration of this observation, we calculate $J_{1lin}$ explicitly in
the case of de Sitter space.  We need to obtain $\mid \varphi^H_k(t) \mid^2$
in order to evaluate the integral in (\ref{linearT}). Inserting the term
proportional to
$T_o^2$ in the Hartree equations, we find that $\mid \varphi^H_k(t) \mid^2$
obeys the
differential equation
\begin{equation}
\left[ \frac{d^2}{dt^2}+3H\frac{d}{dt}+\left[\frac{k^2}{a_o^2}+m^2(T_o)+
\frac{\lambda_R T^2_c}{24}\right]e^{-2Ht}-\frac{\lambda_R T^2_c}{24}
\right]\varphi^H_k(t) = 0 \label{desit}
\end{equation}
The solution of this equation is given by:
\begin{eqnarray}
\varphi^H_k(t) & = & \left[ C_1 H^{(1)}_{3/2}(B_ke^{-Ht})+
C_2 H^{(2)}_{3/2}(B_ke^{-Ht}) \right] e^{-\frac{3}{2}Ht}
\label{sols} \\
     B_k & = & \frac{1}{H}\left[\frac{k^2}{a_o^2}+m^2(T_o)+
\frac{\lambda_R T^2_c}{24}\right]^{\frac{1}{2}} \nonumber
\end{eqnarray}
where $H_{3/2}^{(1,2)}$ are the Hankel functions and we have assumed
\[m^2(T_o)\; ; \; \frac{\lambda_R T^2_c}{24} \ll H . \]

The coefficients $C_1 , C_2$ are determined by the initial conditions
on $\varphi^H_k(t)$ described above.  We finally obtain
\begin{equation}
J_{1lin} = \frac{m(T_o)T_o}{8\pi^2}\left(H e^{Ht_o}\right)^4
\int^{\infty}_{1}
\frac{dz}{z} \frac{\sqrt{z^2-1}}{\left[\frac{\lambda_R T^2_c}{24}+m^2(T)z
\right]^2}
\end{equation}
This term is time independent, finite and positive. This example clearly
illustrates the fact that different powers of $T_o$ enter in the expansion
with different functions of time and that the high temperature expansion
is non-uniform as a function of time.

\section{Evolution of the Initial Distribution}

The initial density matrix at $t=t_o$ was assumed to be thermal for the
adiabatic modes. This corresponds to a Boltzmann distribution for the
uncoupled harmonic oscillators for the adiabatic modes of momentum $\vec{k}$,
and frequencies ${\cal{W}}_k(t_o)$ (in the Hartree approximation; as usual,
the one-loop result can be found by replacing this with $W_k(t_o)$). That is
\begin{eqnarray}
\rho(t_o) & = &  \frac{e^{-\beta_o H_o}}{Tr e^{-\beta_o H_o}}
\label{diagdensmat} \\
H_o       & = & \sum_{\vec{k}}\hbar {\cal{W}}_k(t_o)\left[
\alpha^{\dagger}_k(t_o)\alpha_k(t_o)+\frac{1}{2}\right] \label{iniham}
\end{eqnarray}
\noindent The creation and annihilation operators  define the initial
occupation
number of the adiabatic modes:
\begin{eqnarray}
\hat{N}_k(t_o)                 & = &
\alpha^{\dagger}_k(t_o)\alpha_k(t_o) \label{numberop}\\
\langle \hat{N}_k(t_o) \rangle & = &
\frac{1}{e^{\beta_o \hbar {\cal{W}}_k(t_o)}-1} \label{expecnum},
\end{eqnarray}
where the expectation value in (\ref{expecnum}) is in the initial density
matrix at time $t_o$.

In a time dependent gravitational background the concept of particle is
ill-defined. However, by postulating an equilibrium initial density matrix of
the above form, a preferred ``pointer'' basis is singled out at the initial
time. It is this basis that provides a natural definition of particles  at the
initial time and we can use it to ask: how does the expectation value of {\it
this} number operator evolve in time?

At any time $t$, this expectation value is given by
\begin{equation}
\langle \hat{N}_k \rangle (t) = \frac{Tr
\alpha^{\dagger}_k(t_o)\alpha_k(t_o) \rho(t)}{Tr \rho(t_o)}
\label{timedepnum}
\end{equation}
This quantity gives information on how the {\em original} Boltzmann
distribution function for the adiabatic modes evolves with time. The $\vec{k} =
0$
mode will receive a contribution from the order parameter, but since the number
of particles is not conserved (no charge) there is no bose condensation and the
$\vec{k} =0$ mode will give a negligible contribution to the total number of
particles. Thus we only concentrate on the $\vec{k} \neq 0$ modes.

The expectation value (\ref{timedepnum}) may be easily computed by writing the
creation and annihilation operators in terms of $\eta_k,\ \Pi_k= \delta \slash
\delta \eta_{-k}$ in the {Schr\"{o}dinger} picture at $t_o$. The result of
doing
this is:

\begin{eqnarray}
\alpha^{\dagger}_k(t_o) & = & \frac{1}{\sqrt{2\hbar}}\left[
-\frac{1}{\sqrt{a^3(t_o){\cal{W}}_k(t_o)}}\frac{\delta}{\delta \eta_k}
+\sqrt{a^3(t_o){\cal{W}}_k(t_o)}\eta_{-k} \right] \label{crea} \\
\alpha_k(t_o)           & = & \frac{1}{\sqrt{2\hbar}}\left[
\frac{1}{\sqrt{a^3(t_o){\cal{W}}_k(t_o)}}\frac{\delta}{\delta \eta_{-k}}
+\sqrt{a^3(t_o){\cal{W}}_k(t_o)}\eta_{k} \right] \label{dest}
\end{eqnarray}

After some straightforward algebra we find
\begin{eqnarray}
\langle \hat{N}_k \rangle (t)+ \frac{1}{2} & = &
\left(2 \mid {\cal{F}}_k(t,t_o) \mid^2 -1 \right)
\left( \langle \hat{N}_k \rangle (t_o)+ \frac{1}{2} \right)
\label{numrel} \\
\mid {\cal{F}}_k(t,t_o) \mid^2             & = &
\frac{1}{4}\frac{\mid \varphi^H_k(t) \mid^2}{\mid \varphi^H_k(t) \mid^2o}
 \left[1+ \frac{a^6(t)}{a^6(t_o){\cal{W}}^2_k(t_o)}
\frac{\mid{\dot{\varphi}}^H_k(t)\mid^2}{\mid \varphi^H_k(t) \mid^2}\right] +
\frac{1}{2}
 \label{bogol}
\end{eqnarray}
where we have made use of (\ref{arealfin}, \ref{aimagfin}). This result
exhibits the two contributions from ``spontaneous'' (proportional to the
initial thermal occupation) and ``induced'' (independent of it).

We now show that this result may be understood as a Bogoliubov
transformation. To do this, consider the expansion of the field in the
{\it Heisenberg} picture:
\begin{equation}
\eta_k(t) = \frac{1}{\sqrt{2}} \left( \tilde{\alpha}_k
 \varphi^{H\dagger}_k(t)+
{\tilde{\alpha}}^{\dagger}_k \varphi^H_k(t) \right) \label{heisfield}
\end{equation}
where the mode functions satisfy
\begin{equation}
\frac{d^2 \varphi^H_k}{dt^2} + 3\frac{\dot{a}}{a}\frac{d\varphi^H_k}{dt}+
\left[\frac{\vec{k}^2}{a^2}+{\cal{V}}^{(2)}(\phi(t))\right]\varphi^H_k =0
\label{heishart}
\end{equation}
together with the self-consistency relation.
Then the Heisenberg field $\eta_k(t)$ is a solution of the Hartree
Heisenberg equations of motion and the $\tilde{\alpha}^{\dagger}_k \; ,
\tilde{\alpha}_k$ create and destroy the Hartree-Fock states. Notice
that in
the Heisenberg picture, these creation and annihilation operators {\it
do not depend on time}. Using the Wronskian properties of the functions
$\varphi^H_k(t)$ (see appendix A) we can invert and find the creation and
annihilation operators in terms of $\eta_k(t)$ and its canonically
 conjugate momentum $\Pi_{-k}(t)$. Once we have expressed
 these operators in the
 Heisenberg picture in terms of the field and its canonically conjugate
momentum, we can go to the {Schr\"{o}dinger} picture at time $t_o$.
In this picture the creation and annihilation operators
 depend on time and are given by
\begin{eqnarray}
\tilde{\alpha}_k (t) & = & \frac{i}{\sqrt{2}}\left[\Pi_k(t_o)
\varphi^H_k(t)-a^3(t)\eta_k(t_o)\dot{\varphi}^H_k(t) \right]
 \label{schpicdag} \\
{\tilde{\alpha}}^{\dagger}_{-k} (t)
      & = &  \frac{-i}{\sqrt{2}}\left[\Pi_k(t_o)
\varphi^{H\dagger}_k(t)-a^3(t)\eta_k(t_o)\dot{\varphi}^{H\dagger}_k(t)\right]
 \label{schpic}
\end{eqnarray}

The {Schr\"{o}dinger} picture fields at $t_o$ can be written in terms of the
operators (\ref{crea}, \ref{dest}) and we finally find the creation and
destruction operators at time $t$ to be related to those at time $t_o$
by a Bogoliubov transformation:
\begin{equation}
\tilde{\alpha}_k (t) = {\cal{F}}_{+,k}(t,t_o) {\alpha}_k(t_o) +
{\cal{F}}_{-,k}(t,t_o)
{\alpha}^{\dagger}_{-k} (t_o) \label{bogoltrans}
\end{equation}
If we now compute the average of the new creation and annihilation operators in
the initial density matrix and write the mode functions $\varphi_k(t)$ in terms
of the real functions ${\cal{U}}_{1,2}$ as defined in  appendix A, we recognize
$\mid {\cal{F}}_{+,k}(t,t_o) \mid^2 $ to be the same as $\mid
{\cal{F}}_{k}(t,t_o) \mid^2$ given by (\ref{bogol}).

We also find that
\[\mid {\cal{F}}_{+,k}(t,t_o) \mid^2 -
 \mid {\cal{F}}_{-,k}(t,t_o) \mid^2 = 1 \]
as is required for a Bogoliubov transformation.

One way to interpret this result is that, at least within the one-loop or
Hartree approximations, time evolution corresponds to a Bogoliubov
transformation. This interpretation is, in fact, consistent with the result
that
in these approximation schemes, the density matrix remains Gaussian with the
only change being that the covariance and mixing terms change with time.

Thus within the one-loop or Hartree approximation, time evolution corresponds
to a ``squeezing'' of the initial state. The covariance changes with time and
this corresponds to a Bogoliubov transformation.
As argued by Grishchuk and Sidorov\cite{grishchuk} the amplification of quantum
fluctuations  during inflation is a process of quantum squeezing and it
corresponds to a Bogoliubov transformation.  The properties of these
``squeezed'' quantum states have been investigated in references
\cite{branden,gasperini,albrecht}.

Hu and Pavon\cite{hupavon}, Hu and Kandrup\cite{hukandrup} and
Kandrup\cite{kandrup} have introduced a non-equilibrium, coarse-grained entropy
that grows in time as a consequence of particle production and ``parametric
amplification''. This definition was generalized by Brandenberger et
al. \cite{branden}, and Gasperini and  collaborators \cite{gasperini} to give a
measure of the entropy of the gravitational field. The growth of this entropy
is again a consequence of the parametric amplification of fluctuations and the
``squeezing'' of the quantum state under time evolution.

These authors argue that the non-equilibrium coarse-grained entropy in the
mode of (comoving) wavevector $\vec{k}$ is
\begin{equation}
s_k \approx \ln\left(\langle N_k \rangle (t)\right)
\label{entropy}
\end{equation}
in the case when $\langle N_k \rangle (t) \gg 1$.
Thus the growth of entropy is directly associated with ``particle
production'' or in our case to the evolution of the initial Boltzmann
distribution function.

The coefficient of parametric amplification is related to the Bogoliubov
coefficient  given by equation
(\ref{bogol}). Thus this coefficient directly determines the time dependence
of the non-equilibrium coarse-grained entropy.

\section{Two simple examples:}

As an example of the method that allows the out-of equilibrium time
evolution, we solve numerically the simple cases of a free massive scalar
field in two relevant cosmologies:

\subsection{de Sitter cosmology:}

In this case the scale factor is $a(t)=a_o e^{Ht}$ with H being Hubble's
constant. The important quantities that encode the out of
equilibrium  evolution are the mode functions $\varphi_k(t)$ that obey the
equation (\ref{fievolution}) with $V''(\phi)=m^2$. With these mode
functions we construct the real and imaginary parts of $A_k(t)$ and  the
kernels of the density matrix (see eq. \ref{Aoftime}) and
the parametric amplification factor (\ref{bogol}). It is convenient to
rescale the differential equation and functions and define
\begin{equation}
z = H(t-t_o) \; ; \; q = \frac{k e^{-Ht_o}}{a_o H} \; ; \;
\epsilon = \frac{m}{H}
\end{equation}
the variable $q$ is recognized as the physical wavevector at the initial
time $t_o$ multiplied by the horizon size (or equivalently, horizon size
divided by physical wavelength).
At the same time it is convenient to rescale the mode functions
(in terms of their real
and imaginary parts)
\begin{eqnarray}
\Psi^{R,I}_q(z)                      & = &  (a_o^3H)^{\frac{1}{2}}
\varphi^{R,I}_k(t)
\label{rescfi} \\
\Psi^{R}_q(0)                        & = &
\left[q^2+\epsilon^2\right]^{-\frac{1}{4}}
\; \; ; \; \; \Psi^{I}_q(0) = 0 \label{rescboun1}\\
\frac{d \Psi^{R}_q(z)}{dz}\mid_{z=0} & = & 0 \; \; ; \; \;
\frac{d \Psi^{I}_q(z)}{dz}\mid_{z=0} =
\left[q^2+\epsilon^2\right]^{\frac{1}{4}} \label{rescboun2}
\end{eqnarray}

The differential equation becomes
\begin{equation}
\left[\frac{d^2}{dz^2}+3\frac{d}{dz}+q^2 e^{-2z}+\epsilon^2 \right]
 \Psi^{R,I}_q(z)=0
 \end{equation}
 We have numerically integrated these equations with the above boundary
 conditions choosing as representative parameters $m = 1\ {\rm Gev} \; ; \; H =
 10^{10}\ {\rm Gev}$, $q = 0.01,0.1,1,10, 100$. In figure (1.a,b,c) we show
 ${\cal{A}}_{Rk}/(a_o^3H)$ for $q=0.1,1.0,10$ as function of $z$ for $\epsilon=
 10^{-10}$. Figure (2) shows the logarithm of
 ${\cal{A}}_{Ik}e^{-3Ht_o}/(a_o^3H)$ for $q=0.01,0.1,1.0,10,100$ as a function
 of z. Whereas
 the real part tends to (a $q$-dependent) constant at large times, the
 imaginary part grows at long times as $\approx e^{H(t-t_o)}$
 (the slope  on the graph is 1).
 Figure (3)  shows the logarithm of the parametric amplification factor
 (\ref{bogol}) for the
 above values of $q$ as a function of $z$. The slope of the lines at long times
 is 2, thus the parametric amplification factor grows as
 $\approx e^{2H(t-t_o)}$ at large times.
 Thus we clearly see that if the distribution
 function at some initial time $t_o$ was determined by an equilibrium
 Boltzmann factor (in terms of the comoving wavelengths),
 this distribution function evolves in time out of
 equilibrium and grows with time approximately as $\approx e^{2H(t-t_o)}$
 at long times but with
 different rates for different wavevectors.

{}From the numerical integration we see that in the case of
 de Sitter expansion, at late times (when
the number of ``particles produced'' is large and the expression for
the entropy  (\ref{entropy}) is valid) the entropy per mode
 grows {\it linearly}
with comoving time.

\subsection{Radiation dominated cosmology}
For a radiation dominated cosmology the scale factor is given by
\[a(t)=a_o \left(\frac{t}{t_o}\right)^{\frac{1}{2}} \]
For this case, a convenient rescaling is in terms of the variables

\[ s=mt \; \; ; \; \; q = \frac{k}{m a_o} \; \; ; \; \;
\Psi_q(s) = \left(a_o^3 m \right)^{\frac{1}{2}} \varphi_k(t) \]

The differential equation for the mode functions, and boundary conditions
thus become
\begin{equation}
\left[\frac{d^2}{ds^2}+\frac{3}{2s}\frac{d}{ds}+\frac{q^2}{s}+1 \right]
 \Psi^{R,I}_q(s)=0 \label{psiradom}
\end{equation}
\begin{eqnarray}
\Psi^{R}_q(s_o)                        & = &
\left[q^2+1 \right]^{-\frac{1}{4}}
\; \; ; \; \; \Psi^{I}_q(s_o) = 0 \label{radomboun1}\\
\frac{d \Psi^{R}_q(s)}{ds}\mid_{s=s_o} & = & 0 \; \; ; \; \;
\frac{d \Psi^{I}_q(s)}{ds}\mid_{s=s_o} =
\left[q^2+1 \right]^{\frac{1}{4}} \label{radomboun2}
\end{eqnarray}

There are several noteworthy features in the  radiation dominated case as
exhibited by figures (4.a-d) and (5.a,b). There is a strong dependence on
the initial condition as parametrized by $t_o$ ($s_o=mt_o$) and also a
strong dependence on the initial physical wavenumber $q$. Perhaps the
most notable feature are the oscillations in both the real and imaginary
parts of the covariance ${\cal{A}}$ that translate into an oscillatory
behavior in the parametric amplification factor. As result, at long times
the entropy per mode (\ref{entropy}) {\it is not} a monotonically increasing
function of comoving time.

\section{\bf Conclusions}

Non-equilibrium aspects of the dynamics of scalar fields in spatially
flat FRW cosmologies were studied by means of a functional {Schr\"{o}dinger}
approach. The initial state was specified as a thermal density matrix at
some early initial time assuming local thermodynamic equilibrium for the
adiabatic modes at that particular time. This density matrix was evolved
in time and the evolution equations for the order parameter (ensemble
average of the scalar field) and the fluctuations were obtained both
to one-loop and in a non-perturbative self-consistent Hartree approximation.
The renormalization aspects were studied in detail and it was pointed out
that the renormalization counterterms contain a dependence on the initial
conditions through the scale factor and its derivatives at the initial
time.

The high temperature expansion was investigated and it was found that the
limit of high temperatures and long times are not interchangeable. As a
consequence of the red-shift of the initial temperature the coefficients
of the different powers of temperature are different time-dependent
functions. The high
temperature expansion is only valid within a short time interval after the
initial time and certainly breaks down at long times.

The time evolution of the Boltzmann distribution functions (initially the
thermal equilibrium distribution functions) is obtained. It is pointed out
that to one-loop order and also in the Hartree approximation, the time
evolved density matrix describes quantum ``squeezed'' states and time
evolution corresponds to a Bogoliubov transformation.

To illustrate the departure of equilibrium, we have studied numerically the
case of a free massive scalar field in de Sitter and radiation dominated
cosmologies. It was found that a suitably defined coarse-grained
non-equilibrium entropy (per $\vec{k}$ mode) grows linearly with time in
the de Sitter case but it is not a monotonically increasing function of
time in the radiation dominated case. This result may cast some doubt on the
applicability of this definition of the non-equilibrium entropy.
There still remain some (open) fundamental questions regarding the
connection of this entropy and the thermodynamic entropy of the universe,
in particular whether the amount of entropy produced is consistent with
the current bounds.

This work sets the stage for a numerical study of the dynamics of phase
transitions in cosmology fully incorporating the non-equilibrium aspects in the
evolution of the order parameter and which at the same time can account for the
dynamics of  the fluctuations which will necesarily become very important
during the phase transition.

We expect to report on the numerical study of the phase transition in a
forthcoming article \cite{us}.

\acknowledgements

D. B. would like to thank G. Veneziano, M. Gasperini,
R. Brandenberger, E. Copeland, and M. Hindmarsh,
for illuminating conversations and remarks, the
Laboratoire de Physique Th\'eorique at Universit\'e de Paris VI
for hospitality and
N.S.F. for support through grant: PHY-9302534. He would also like to thank the
Institute for Theoretical Physics for hospitality and partial support through
N.S.F grant: PHY-89-04035  during part of this work.  This work was
partially supported by a France-U.S.A.
 binational collaboration between CNRS and N.S.F.
through N.S.F. grant No: INT-9216755.
H. J. de V. wishes to thank the participants of the Imperial-Cambridge-Sussex
seminars for stimulating remarks and the Department of Physics and
Astronomy, University of Pittsburgh, for hospitality.
R.H. was supported in part by D.O.E. contract $\#$ DOE-ER/408682.

\vspace{36 pt}

\appendix
\section{}
The Riccati equation (\ref{riccati}) can be transformed into a linear
differential equation by the change of variables
\begin{equation}
{\cal{A}}_k(t) = -ia^3(t)\frac{\dot{\varphi}_k(t)}{\varphi_k(t)}
\label{fiofk}
\end{equation}
We find that $\varphi_k(t)$ obeys a simple evolution equation
\begin{equation}
\frac{d^2 \varphi_k}{dt^2} + 3\frac{\dot{a}}{a}\frac{d\varphi_k}{dt}+
\left[\frac{\vec{k}^2}{a^2}+V''(\phi_{cl}(t))\right]\varphi_k =0
\label{fievolution}
\end{equation}
with $\phi_{cl}(t)$ the solution to the {\it classical} equation of
motion (\ref{pieq}, \ref{fieq}).
In the Hartree approximation $V''(\phi_{cl}(t))$ should be replaced
by ${\cal{V}}^{(2)}(\phi(t))$ with $\phi(t)$ the {\it full solution}
of the self-consistent equations.
\begin{equation}
\frac{d^2 \varphi^H_k}{dt^2} + 3\frac{\dot{a}}{a}\frac{d\varphi^H_k}{dt}+
\left[\frac{\vec{k}^2}{a^2}+{\cal{V}}^{(2)}(\phi(t))\right]\varphi^H_k =0
\label{fihevolution}
\end{equation}

The Wronskian for two arbitrary solutions to the above differential
equation is
\begin{equation}
{\bf{W}}[\varphi_1,\varphi_2] = \dot{\varphi_2}(t)\varphi_1(t)
-\dot{\varphi_1}(t)\varphi_2(t) = \frac{C}{a^3(t)} \label{wronskian}
\end{equation}
with $C$ a constant. Then writing $\varphi_k(t) = \varphi_{1k}(t)+i
\varphi_{2k}(t)$ with $\varphi_{1,2}$ {\it real solutions} we find
\begin{eqnarray}
{\cal{A}}_{Rk}(t) & = & \frac{C}{\left[\varphi^2_{1k}+\varphi^2_{2k}\right]}
\label{area} \\
{\cal{A}}_{Ik}(t) & = & -a^3(t)\left[\frac{\dot{\varphi_{1k}}\varphi_{1k}+
\dot{\varphi_{2k}}\varphi_{2k}}{\varphi^2_{1k}+\varphi^2_{2k}}\right]
\label{aima}
\end{eqnarray}

 It proves convenient to introduce (for all $\vec{k}$)
 the real functions ${\cal{U}}_{1,2}(t)$
as
\begin{equation}
\varphi_{1,2}(t) = [a(t)]^{-\frac{3}{2}}
\frac{{\cal{U}}_{1,2}(t)}{\sqrt{W_k(t_o)}}
 \label{calU}
\end{equation}
The ${\cal{U}}_{\alpha k}$ for $\alpha=1,2$ are real and
satisfy the {Schr\"{o}dinger}-like differential equation
\begin{eqnarray}
& & \left[\frac{d^2}{dt^2}-\frac{3}{2}\left(\frac{\ddot{a}}{a}+\frac{1}{2}
\frac{\dot{a}^2}{a^2}\right)+\frac{\vec{k}^2}{a^2(t)}+
V''(\phi_{cl})(t)\right]
{\cal{U}}_{\alpha k}(t)
=0 \label{diffeqU} \\
& & {\bf{W}}[{\cal{U}}_{1,k},{\cal{U}}_{2,k}]= C W_k(t_o)
\label{wronskchi}
\end{eqnarray}
with ${\bf{W}}[\cdots]$ the Wronskian, and $C$ is
the same constant as above.
Since the choice of $C$ corresponds to a choice of normalization of
these functions, we choose $C=1$. The initial conditions (\ref{Are},
\ref{Aim}) still leave one free condition on these functions, we
choose
\begin{eqnarray}
{\cal{U}}_{1k}(t_o) & \neq 0 & \label{calU1} \\
{\cal{U}}_{2k}(t_o) &  =     & 0 \label{calU2}
\end{eqnarray}

The boundary conditions on the mode functions ${\cal{U}}_{\alpha,k}$
are
\begin{eqnarray}
& & {\cal{U}}_{1k}(t_o) = 1 \; \; ; \; \; {\cal{U}}_{2k}(t_o) = 0
\label{boundconU} \\
& & \dot{{\cal{U}}}_{1k}(t_o) = \frac{3}{2}\frac{\dot{a}(t_o)}{a(t_o)}
\; \; ; \; \; \dot{{\cal{U}}}_{2k}(t_o) = W_k(t_o) =
\left[\frac{\vec{k}^2}{a^2(t_o)}+V''(\phi_{cl}(t_o))\right]^{\frac{1}{2}}
\label{boundconUdot}
\end{eqnarray}
and the corresponding replacement for the Hartree case.
Thus the final solution to the Riccati equation (\ref{riccati}) with
the given initial conditions is
\begin{eqnarray}
{\cal{A}}_{Rk}(t) & = & \frac{a^3(t)W_k(t_o)}{\left[{\cal{U}}^2_{1k}(t)+
{\cal{U}}^2_{2k}(t)\right]}
\label{arealfin} \\
{\cal{A}}_{Ik}(t) & = & -a^3(t)\left[ \frac{
{\cal{U}}_{1k}\left(\dot{{\cal{U}}}_{1k}
-\frac{3\dot{a}}{2a}{\cal{U}}_{1k}\right)+
{\cal{U}}_{2k}\left(\dot{{\cal{U}}}_{2k}
-\frac{3\dot{a}}{2a}{\cal{U}}_{2k}\right)}
{ {\cal{U}}^2_{1k}(t)+ {\cal{U}}^2_{2k}(t) }  \right] \label{aimagfin}
\end{eqnarray}

In terms of the original functions $\varphi_k(t)$ (\ref{fiofk}) the
initial conditions are simply
\begin{eqnarray}
\varphi_k(t_o)               & = & \frac{1}{\sqrt{a^3(t_o)W_k(t_o)}}
 \label{bounfi} \\
\dot{\varphi}_k(t)\mid_{t_o} & = &
 i \sqrt{\frac{W_k(t_o)}{a^3(t_o)}}
\label{bounfidot}
\end{eqnarray}

Then the initial conditions for $\varphi_k(t) \; ; \; {\varphi}^*_k(t)$
are naturally interpreted as those for negative and positive (adiabatic)
frequency modes at the initial time $t_o$.

The  reason for introducing the functions ${\cal{U}}_{\alpha,k}(t)$
is because these obey a simpler second order Schr\"{o}dinger-like equation
which is amenable to be studied in the asymptotic regime via WKB
approximations (see the section on renormalization).

\section{Conformal Time Analysis}

It is interesting to see how some of our results can be obtained by rewriting
the metric in terms of the conformal time defined by:
\begin{equation}
\eta = {\int}^{t} \frac{dt'}{a(t')}
\end{equation}

The first thing we should note is that the physics should not depend on what
time coordinate is used, since the theory should be generally coordinate
invariant (there are no gravitational anomalies in four dimensions
\cite{gravanom}). Thus, the field amplitude and the canonical momentum in
conformal time should be related to those in comoving time via a canonical
transformation. We now show that this is indeed the case.

Using the conformal time version of the line element $ds^2 = a^2(\eta)(d\eta^2
- d\vec{x}^2)$, the scalar field action can be written as:

\begin{equation}
S[\Phi] = \int d\eta d^3x a^4(\eta) \left[ \frac{1}{2 a^2(\eta)}
((\partial_{\eta}\Phi)^2 - (\nabla \Phi)^2) -
V(\Phi(\eta, \vec{x}))\right],
\end{equation}
where the potential term is as in the text (i.e. it could include a coupling to
the curvature scalar). Following the standard procedure for obtaining the
canonical momentum $\Pi(\eta, \vec{x})$ to $\Phi(\eta, \vec{x})$, and
to get at the Hamiltonian density yields:

\begin{eqnarray}
\Pi(\eta, \vec{x}) & = & a^2(\eta) \Phi'(\eta, \vec{x})\nonumber \\
{\cal H} & = & \frac{\Pi^2}{2 a^2(\eta)} +
\frac{a^2(\eta)}{2} (\nabla \Phi)
+ a^4(\eta) V(\Phi).
\label{confham}
\end{eqnarray}
Here conformal time derivatives are denoted by a prime.
The generator $H$ of displacements in conformal time is the spatial integral of
${\cal H}$ above. Thus the conformal time Liouville equation reads:

\begin{equation}
i \frac{\partial \rho}{\partial \eta} = [H, \rho].
\end{equation}

If we label the field and its conjugate momemtum in the comoving time frame
(i.e. that of the text) as $\hat{\Phi},\ \hat{\Pi}$ respectively, the
results of section 2 are:

\begin{eqnarray}
\hat{\Pi}(\vec{x},t) & = & a^3(t)\dot{\hat{\Phi}}(\vec{x},t)\nonumber \\
\hat{H} & = &\int d^3x \left\{ \frac{\hat{\Pi}^2}{2a^3}+
\frac{a}{2}(\vec{\nabla}\hat{\Phi})^2+
a^3 V(\hat{\Phi}) \right\}
\end{eqnarray}

The Liouville equation in comoving time can be rewritten in conformal time
using the relation: $\partial \slash {\partial t} = a(\eta)^{-1}\partial \slash
{\partial \eta}$. After doing this we find that eq.(\ref{liouville}) becomes:
\begin{equation}
i\frac{\partial \hat{\rho}(\hat{\Pi}, \hat{\Phi})}{\partial \eta} =
\left[a(\eta) \hat{H}, \hat{\rho}(\hat{\Pi}, \hat{\Phi})\right].
\end{equation}
But
\begin{equation}
a(\eta) \hat{H} = \int d^3x \left[ \frac{\hat{\Pi}}{2 a^2(\eta)} + \frac{1}{2}
a^2(\eta) (\nabla \hat{\Phi})^2 + a^4(\eta) V(\hat{\Phi})\right].
\end{equation}
Comparing this with eq.(\ref{confham}), we see that we can make the
identifications: $\Pi=\hat{\Pi},\ \Psi= \hat{\Psi}$. Thus not only is the
physics equivalent in both coordinate systems (as must have been the case), but
the physics in the two coordinate systems related by a trivial canonical
transformation.

We can rewrite all of the comoving time results in terms of conformal time.
Some of the equations take on a much simpler form in conformal time than in
comoving time. This will be important when numerical issues are tackled, such
as an analysis of the back-reaction problem. This work is in progress\cite{us}

\newpage

\begin{center}
\underline{\bf Figure Captions:}
\end{center}

\underline{\bf Figure 1(a)}:
${{\cal{A}}_{R}}\slash{a_o^3H}$ vs. z for $q=0.1$, $m/H=10^{-10}$.

\underline{\bf Figure 1(b)}: ${{\cal{A}}_{R}}\slash{a_o^3H}$  vs. z for de
Sitter for $q=1.0$, $m/H=10^{-10}$.

\underline{\bf Figure 1(c)}:
${{\cal{A}}_{R}}\slash{a_o^3H}$  vs. z for de Sitter for $q=10$,
 $m/H=10^{-10}$.

\underline{\bf Figure 2}:
$\ln\left({\cal{A}}_{I}e^{-3Ht_o}/a_o^3H\right)$
 vs. z for  de Sitter for
 $q=0.01;\; \;0.1;\; \; 1; \; \; 10; \; \; 100$, $m/H=10^{-10}$.
The slopes of the lines at long times is 1.

\underline{\bf Figure 3}:
$\ln \mid {\cal{F}} \mid$
 vs. z for de Sitter for
 $q=0.01;\; \;0.1;\; \; 1; \; \; 10; \; \; 100$, $m/H=10^{-10}$.
The slope of the lines at long times is 2.

\underline{\bf Figure 4(a)}:
${{\cal{A}}_{R}}\slash{a_o^3m}$ vs. $s$ for radiation dominated for $q=1$
(solid
   line)
and $q=10$ (dashed line); $s_o=1.0$.

\underline{\bf Figure 4(b)}: ${{\cal{A}}_{I}}\slash{a_o^3m}$ vs. $s$ for
radiation dominated for $q=1$ (solid line) and $q=10$ (dashed line);
$s_o=1$.

\underline{\bf Figure 4(c)}: ${{\cal{A}}_{R}}\slash{a_o^3m}$ vs. $s$ for
radiation dominated for $q=1$ (solid line) and $q=10$ (dashed line);
$s_o=10$.

\underline{\bf Figure 4(d)}: ${{\cal{A}}_{I}}\slash{a_o^3m}$ vs. $s$ for
radiation dominated for $q=1$ (solid line) and $q=10$ (dashed line);
$s_o=10$.

\underline{\bf Figure 5(a):} $\mid {\cal{F}} \mid^2$ vs. $s$ for radiation
dominated for $q=1$ (solid line) and $q=10$ (dashed line); $s_o=1$

\underline{\bf Figure 5(b):}
$\mid {\cal{F}} \mid^2$ vs. $s$ for radiation dominated for $q=1$ (solid
line) and $q=10$ (dashed line); $s_o=10$

\end{document}